# A Risk Profile for Information Fusion Algorithms

**Kenric P. Nelson [1],\*, Brian J. Scannell [1] and Herbert Landau [2]**

[1] Raytheon Integrated Defense Systems, 235 Presidential Way, Woburn, MA 01801, USA;
E-Mail: Brian_J_Scannell@raytheon.com

[2] Raytheon Integrated Defense Systems, 2461 S. Clark St., Suite 1000, Arlington, VA 22202, USA;
E-Mail: Herbert_Landau@raytheon.com

\* Author to whom correspondence should be addressed; E-Mail: Kenric_P_Nelson@raytheon.com.

**Abstract:** E.T. Jaynes, originator of the maximum entropy interpretation of statistical mechanics, emphasized that there is an inevitable trade-off between the conflicting requirements of robustness and accuracy for any inferencing algorithm. This is because robustness requires discarding of information in order to reduce the sensitivity to outliers. The principal of nonlinear statistical coupling, which is an interpretation of the Tsallis entropy generalization, can be used to quantify this trade-off. The coupled-surprisal, $-\ln_\kappa(p) \equiv -\frac{p^\kappa - 1}{\kappa}$, is a generalization of Shannon surprisal or the logarithmic scoring rule, given a forecast $p$ of a true event by an inferencing algorithm. The coupling parameter $\kappa = 1 - q$, where $q$ is the Tsallis entropy index, is the degree of nonlinear coupling between statistical states. Positive (negative) values of nonlinear coupling decrease (increase) the surprisal information metric and thereby biases the risk in favor of decisive (robust) algorithms relative to the Shannon surprisal ($\kappa = 0$). We show that translating the average coupled-surprisal to an effective probability is equivalent to using the generalized mean of the true event probabilities as a scoring rule. The metric is used to assess the robustness, accuracy, and decisiveness of a fusion algorithm. We use a two-parameter fusion algorithm to combine input probabilities from N sources. The generalized mean parameter 'alpha' varies the degree of smoothing and raising to a power $N^\beta$ with $β$ between 0 and 1 provides a model of correlation.

## 1. Introduction

The assessment techniques used to evaluate the performance of information fusion and inferencing algorithms have a significant influence in the selection of optimal algorithms. The foundational metrics are misclassification rate and the Shannon and Brier proper scoring rules [1–3]. However, these metrics do not fully capture the trade-off between the need for fusion algorithms to make decisions, provide accurate inferences in the form of probabilities, and be robust against anomalies [4]. For an algorithm which classifies based on a threshold probability, the classification cost function is simply a step-function at the threshold. Closely related to classification performance is the Brier or mean-square error, which provides a cost function based on a uniform likelihood of the decision boundary [5]. However, the mean-square error does not fully capture the meaning of an accurate probability which is bounded by impossible ($p = 0$) and certain ($p = 1$) events. The Shannon surprisal or logarithmic average of the inverse probability accurately reflects the infinite cost of falsely reporting an impossible event. Nevertheless, the surprisal score is in practice often viewed as oversensitive to individual test samples, in assessing the average performance of an inference engine. There is a need for an assessment methodology which balances the requirements for accuracy in the reported probabilities, with robustness to errors, and performance of classification decisions.

By demonstrating the equivalence between maximizing the generalized mean of the reported probability of a true event, and minimizing a cost function based on the non-additive entropy [6–9] originally proposed by C. Tsallis, we have developed a simple methodology to assess the decisiveness, accuracy, and robustness of fusion and inferencing algorithms. The equivalence is facilitated by a physical interpretation of Tsallis entropy reflecting the degree of nonlinear coupling between the statistical states of a system [10]. In this view, Shannon information theory is based on the assumption of mutually exclusive statistical states, while non-additive entropy models systems in which nonlinearity creates strong dependence between the statistical states. The surprisal cost function is equivalent to the geometric mean of the reported probabilities, *i.e.*, the joint probability is the multiplication of the independent probabilities. In assessing inferencing algorithms, the level of risk tolerance is modeled as a source of nonlinear coupling. By varying the degree of nonlinear coupling, 'Decisive' and 'Robust' risk bounds on the 'Accuracy' or 'Neutral' risk reflected by the average surprisal are established with positive and negative coupling coefficients for the 'coupled-Surprisal'. However, the average 'coupled-Surprisal' is equivalent to fusing the reported probabilities using the generalized mean. The three effective probabilities, which we refer to as the 'Decisive Probability', 'Accurate or Risk Neutral Probability' and 'Robust Probability' provide a risk profile of the inference algorithm.

The remainder of this paper is divided as follows: Section 2 provides background on a fusion algorithm with two power-law parameters and the concept of nonlinear statistical coupling. The alpha parameter forms the generalized mean and the beta parameter determines the effective independence. Section 2.2 reviews non-additive entropy and nonlinear statistical coupling as a model for risk. Section 3 shows that the generalized mean is the probability combining rule for the generalized entropies defined by Renyi and Tsallis. In Section 4 an example demonstrating the application of the generalized mean for designing and assessing a fusion algorithm is illustrated. A conclusion and suggestions for future investigation are provided in Section 5.

## 2. Background

*2.1. Alpha-Beta Fusion Algorithm*

As the complexity of information systems has increased, methods which can model the errors and correlations in fusion of large sources of information while minimizing the growth of the computational complexity have become important. While Bayesian analysis provides a theoretical foundation for the fusion of new information, modeling the pair-wise correlations between information sources grows quadratically with the number of nodes. Research in the statistical physics community has shown that long-range correlations within complex systems can be modeled by raising probabilities to a power. The *q*-calculus and *q*-statistics [11,12] which developed from this analysis provides a theoretical foundation for modeling correlations with power-law relations. The *q*-product, a deformed product rule, which has been used to generalize Gauss' law of errors [13,14] and the central-limit theorem [15], provides a method of fusing probabilities with a nonlinear dependency structure. In the machine learning community, investigators have demonstrated the effectiveness of averaging probabilities as a simple fusion method [16–18] which is more robust to errors and approximates the correlations between inputs. This provides a simple, effective alternative to the naïve Bayes method of multiplying probabilities. Combining probabilities using the generalized mean [19,20] provides flexibility in the degree of smoothing, but is only a limited model of system wide correlations.

By combining the generalized mean, with an additional power term $N^\beta$ representing the effective number of independent samples, both the need to smooth errors and account for correlations is modeled. The probability of class $C_i$, conditioned on multiple data inputs, $X_i$ is approximated by [21]:

$$P(C_i \mid X_1 = x_1, ... X_N = x_N) \approx \frac{1}{Z} \left( \sum_{i=1}^{N} w_i P_i^\alpha (x_i \mid C_i) \right)^{\frac{W^\beta}{\alpha}} P(C_i); \quad W = \sum_{i=1}^{N} w_i \qquad (1)$$

where Z is the normalization over the classes and $w_i$ is a weighting of the input probabilities. The generalized mean of the likelihood function $P(x_i/C_i)$ is computed where $\alpha$ is the parameter of the generalized mean. The prior probability $P(C_i)$ is considered independent of the likelihoods. $\beta$ ranges from 0 (fully correlated likelihoods) to 1 (fully independent likelihoods). Combinations of alpha and beta provide both well-known combining rules such as the naïve-Bayes ($\alpha = 0$, $\beta = 1$), log-average (0,0), average (1,0) and a continuum of combinations between these rules which provide flexibility in modeling error and correlation. The weighting parameter can be used for confidence measures on the individual inputs. While these weights will not be examined here, use of the assessment tools described here on individual inference algorithms could provide a method for assigning these weights.

The design of cost functions for training and assessment of the alpha-beta parameters is critical for the fusion algorithm to be effective. To this end we compare two well known scoring rules, surprisal and Brier, and propose a new score which incorporates a model of risk. The surprisal scoring rule is the information metric of Shannon entropy, $-\ln p_{true,i}$, where $p_{true,i}$ is the probability of the true class for the $i^{th}$ sample. While this metric accurately reflects the cost of information for a probability, the asymptotic approach to infinite as the probability approaches zero, is a severe cost which has been criticized as not reflective of the performance of forecasting systems. Within forecasting a common alternative is the Brier or mean-square scoring rule, $(1 - p_{true,i})^2$, which limits the cost of reporting zero probability to 1. Like surprisal, the Brier is a proper score, which requires that the reported forecast is unbiased relative to the expected probability. Nevertheless, an overreliance on the mean-square average, which does not reflect the full characteristics of an information metric, can encourage the design of inferencing algorithms which allow very confident probabilities. This over-confidence reduces the robustness of the algorithm. This can be examined by considering a direct model of risk and robustness using the principals of nonlinear statistical coupling and will be evident in the example shown in Section 4.

*2.2. Modeling Risk with Coupled-Surprisal*

Topsoe, Anteneodo, and other investigators [22–24] have shown that non-additive entropy provides a model of apriori risk or belief for information systems. We will simplify and expand upon this model using the degree of nonlinear statistical coupling κ to model the negative risk or optimism. We begin with a review of non-additive entropy and show how the requirements for a consistent model of risk provide evidence for a particular form of this model. The field of nonextensive statistical mechanics developed from efforts to provide a thermodynamic model of complex systems influenced by nonlinear dynamics. The key insight proposed by Tsallis [25] was that deforming the probability distributions by a power $p_i^q$ led to an entropy function $\frac{1}{1-q}(-1+\sum_{i=1}^{N} p_i^q)$ which is non-additive for independent systems $H_q(AB) = H_q(A) + H_q(B) + (1-q)H_q(A)H_q(B)$. The maximum entropy distribution for non-additive entropy constrained by the deformed probabilities is referred to as a *q*-exponential or *q*-Gaussian $p_i \propto (1-(1-q)x^2)^{\frac{1}{1-q}}$. This methodology has been shown to provide an effective model of the origin of power-law phenomena in a variety of complex system [26] such as turbulence, dynamics of solar wind, stock market fluctuations, and biological processes to name just a few. Further theoretical evidence including generalized central limit theorems [15,27], probabilistic models [28–30], and models of statistical fluctuations [31,32] have demonstrated a foundational justification for this approach. Nevertheless, a direct physical interpretation of the symbol *q* has remained elusive. One of the difficulties is that the original convention based on raising probabilities to a power, results in an asymmetry between the mathematical model and the physical principal of non-additivity. By directly modeling the degree of non-additivity as $\kappa = 1 - q$ other physical interpretations are more direct and the mathematical symmetry simplifies the development of consistent models [6,10,33,34]. Kappa is choosen as a symbol because of its usage for coupling coefficients, such as the spring constant and its relation to curvature in geometry. The definition used here is related to Kaniadakis' deformed logarithm $\ln_{\{r,\kappa\}} x \equiv x^r \frac{x^\kappa - x^{-\kappa}}{2\kappa}$ by $\kappa^{coupling} = 2\kappa^{Kaniadakis}$ with $r = \kappa^{Kaniadakis}$ [35,36].

An example of the facility of this parameterization is the escort probability, $\frac{p_i^q}{\sum_{i=1}^{N} p_i^q}$, whose name implies that the deformation does not encode new physical information. However, considering this function in relation to the degree of non-additivity, makes evident that it is the probability of a coupled state, which we will refer to as the 'coupled-probability':

$$P_{\kappa,i} \equiv \frac{p_i^{1-\kappa}}{\sum_{i=1}^{N} p_i^{1-\kappa}} = \frac{p_i \prod_{\substack{j=1 \\ j \neq i, p_j \neq 0}}^{N} p_j^\kappa}{\sum_{i=1}^{N} p_i \prod_{\substack{j=1 \\ j \neq i}}^{N} p_j^\kappa} \quad (2)$$

So a defining model of κ is the degree of nonlinear coupling between the statistical states of a system or simply the 'nonlinear statistical coupling' [10]. This connects the parameter directly with the original motivation to model the statistics of nonlinear systems. In [37] the generalized moments where shown to be related to power factors of $(n+1)-nq$. Again the asymmetry obscures the physical interpretation, which is simplified by the expression for 'coupled-moments':

$$\langle x^n \rangle_{n\kappa} \equiv \sum_{i=1}^{N} x^n \frac{p_i^{1-n\kappa}}{\sum_{i=1}^{N} p_i^{1-n\kappa}} \tag{3}$$

This clarifies that the deformed moments require that the coupling parameter $\kappa$ be scaled by the degree of the moment. The coupled-moments are used as constraints in determining the maximum coupled-entropy distribution.

Whereas Shannon entropy provides a metric for information, we will use the non-additive entropy or coupled-entropy to model the biasing of information by risk or belief. This model will provide important insights into the performance of inferencing and fusion algorithms. Using the nonlinear statistical coupling parameter the generalized exponential and its inverse the generalized logarithm are defined as:

$$e_\kappa^x \equiv (1+\kappa x)_+^{1/\kappa}, \quad (y)_+ \equiv \max(0, y)$$
$$\ln_\kappa x \equiv \frac{x^\kappa - 1}{\kappa}, \quad x > 0 \tag{4}$$

In the limit as $\kappa \to 0$ these generalizations converge to the exponential and natural logarithm functions. From the coupled-logarithm we see that $\kappa$ specifies the power of the nonlinear system being modeled. The operation of raising to a power modifies both the argument and the coupling parameter:

$$\left(e_\kappa^x\right)^a = e_{\kappa/a}^{ax}$$
$$\ln_\kappa x^a = a \ln_{a\kappa} x \tag{5}$$

Tsallis entropy can be expressed using the coupled-logarithm in three equivalent forms:

$$H_\kappa(p) = \sum_i p_i \ln_\kappa \left(\frac{1}{p_i}\right) = -\sum_i p_i \ln_{-\kappa}(p_i) = -\sum_i p_i^{1-\kappa} \ln_\kappa(p_i) \tag{6}$$

These forms create some ambiguity regarding the separation of the generalized information metric and the averaging of the information. The first two forms use standard averaging and an information metric of $\ln_\kappa\left(\frac{1}{p_i}\right) = -\ln_{-\kappa}(p_i)$. This form is typically used to define the generalized information form as in [10,22–24]. However, the alternative using the coupled-probability for averaging $p_i^{1-\kappa}$ and an information metric of $-\ln_\kappa(p_i)$, which is shown in Figure 1, has some advantages with regard to a consistent interpretation of $\kappa$ specifying the negative risk or optimism. Positive values of $\kappa$ lower the information cost, which is equivalent to a reduction in risk or an optimistic belief. This domain has a finite cost for probability zero and is associated with the maximum entropy distributions with compact-support domain, which have a finite domain of non-zero probabilities. The cost of information for negative values of $\kappa$ approaches infinity faster than the Shannon surprisal. This domain is associated with the heavy-tail maximum entropy distributions, and is consistent with the higher cost of information selecting distributions which are more 'robust' in that they model a slower decay to states which can be ignored.

**Figure 1.** Coupled-surprisal cost function, $-\ln_\kappa p_{true}$. The red curve is Shannon surprisal $(\kappa = 0)$; curves above this represent a more robust metric $(\kappa < 0)$; curves below surprisal represent a more decisive metric $(\kappa > 0)$.

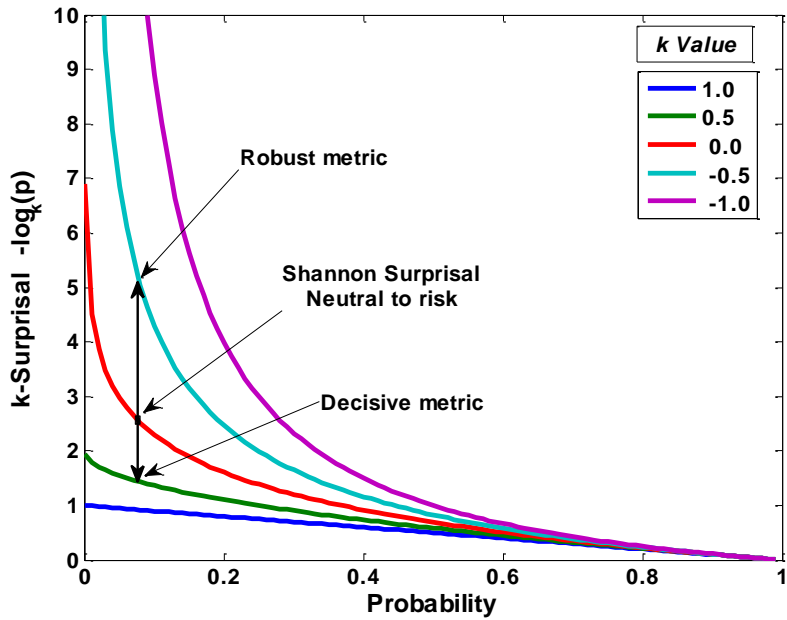

Since the third expression is based on the coupled-moment, renormalization of the entropy function provides additional consistency:

$$H_\kappa^N \equiv -\sum_{i=1}^{N} P_{\kappa,i} \ln_\kappa p_i \tag{7}$$

The normalization of the non-additive entropy has been criticized as not satisfying the Lesche stability criteria [38]. A full discussion of this issue is beyond the scope of this paper; however, as a scoring rule with equally weighted test samples, the coupled-average is not influential. Our primary interest is the relationship between the coupled-surprisal and the generalized mean. The only effect in using the normalized form of the nonextensive entropy is to reverse the sign of $\kappa$ in relation to the generalized mean parameter, which we discuss next.

## 3. Relationship between Generalized Mean and Generalized Entropies

The Tsallis and Renyi entropies can be expressed using the generalized mean:

$$H_\kappa^T = -\ln_{-\kappa}\left(\sum_{i=1}^{N} p_i p_i^{-\kappa}\right)^{-\frac{1}{\kappa}} = \ln_\kappa\left(\sum_{i=1}^{N} p_i p_i^{-\kappa}\right)^{\frac{1}{\kappa}} \tag{8}$$

$$H_\kappa^R = -\ln\left(\sum_{i=1}^{N} p_i p_i^{-\kappa}\right)^{-\frac{1}{\kappa}} \tag{9}$$

The probabilities are separated into $p_i$ and $p_i^{-\kappa}$ to emphasis the relation to the generalized mean with the weight and the sample equal to $p_i$ and the mean parameter $\alpha = -\kappa$. The expression for the Renyi entropy is common with substitution $\kappa = 1-q$; however, the connection between Equations (8) and (6) requires use of the $\kappa$-power and $\kappa$-product:

$$x^{\otimes_\kappa^N} \equiv \prod_{i=1}^{N}{}_{\otimes_\kappa} x \equiv \left(\sum_{i=1}^{N} x^\kappa - (N-1)\right)^{1/\kappa} \tag{10}$$

which following analysis by Oikonomou [39] leads to:

$$\ln_\kappa\left(\sum_{i=1}^{N} p_i p_i^{-\kappa}\right)^{\frac{1}{\kappa}} = \ln_\kappa\left(\prod_{i=1}^{N}{}_{\otimes_\kappa} (1/p_i)^{\otimes_\kappa^{p_i}}\right) = \sum_{i=1}^{N} p_i \ln_\kappa(1/p_i) \tag{11}$$

This structure shows that the difference between the Tsallis and Renyi entropy is the use of $\kappa$-logarithm, which Oikonomou describes as an external correlation distinct from the correlation internal to the distribution formed by the generalized mean.

The relationship between the generalized mean and generalized entropy can be used to define a scoring rule for decision algorithms, which reports an effective probability for the algorithm. The effective probability is determined using the normalized version of Tsallis entropy as the kernel for the coupled-surprisal $-\ln_\kappa p_{i,true}$, with equally weighted test samples $\frac{N^{\kappa-1}}{\sum_{i=1}^{N} N^{\kappa-1}} = \frac{1}{N}$:

$$P_{eff} = \exp_\kappa(-S_\kappa) = \exp_\kappa\left(+\frac{1}{N}\sum_{i=1}^{N} \ln_\kappa p_{true,i}\right)$$

$$= \left(1 - \frac{\kappa N}{\kappa N} + \frac{\kappa}{\kappa N}\sum_{i=1}^{N} p_{true,i}^\kappa\right)^{\frac{1}{\kappa}} = \left(\frac{1}{N}\sum_{i=1}^{N} p_{true,i}^\kappa\right)^{\frac{1}{\kappa}} \tag{12}$$

Again the normalized form is advantageous because there is consistency in the interpretation of $\kappa$ representing the negative risk or optimism, between the maximum entropy distributions, the coupled-surprisal cost function, and the effective probability. The effective probability is a convenient representation of the average score, since test results are commonly expressed as a percentage. The effective probability can be used as a measure of confidence regarding an inferencing algorithm with the different values of $\kappa$ providing insight regarding the effect of risk or other physical models which have a related deformation of the information cost function. This confidence level can serve as a comparison of algorithms and as a weighting of inputs to fusion engines.

Having established the connection between coupled-entropy and the generalized mean, each of the parameters for the alpha-beta fusion method in Equation (1) can be examined in light of the risk bias or degree of nonlinear statistical coupling. Alpha defines the fusion method and is equal to the nonlinear statistical coupling, thus $\kappa_f = \alpha$. The input weights modify the confidence of each input or in terms of risk are $w_i = 1 - \kappa_i$. The output confidence can be split into a portion which is the sum of the input weights $W = \sum_{i=1}^{N} w_i$ and a portion which is the output confidence $W^{\beta-1} = 1 - \kappa_o$. Thus the fusion method can be viewed as controlling the risk bias on the input, fusion, and output of the probabilities:

$$P(C_i \mid X_1 = x_1, \ldots X_N = x_N) \approx \frac{1}{Z}\left( \sum_{i=1}^{N} (1-\kappa_i) P_i^{\kappa_f}(x_i \mid C_i) \right)^{\frac{W(1-\kappa_o)}{\kappa_f}} P(C_i) \tag{13}$$

The relationship between input and output confidence weights and the nonlinear statistical coupling term is seen more clearly using the coupled-logarithm function. Since the generalized mean includes the sum of coupled-logarithms, consider how the inputs weights affect this sum

$$\left[ \exp_{\kappa_f \atop W} \left[ \sum_{i=1}^{N} (1-\kappa_i) \log_{\kappa_f} P_i(x_i \mid C_i) \right]\right]^{1-\kappa_0} = \left[ \exp_{\kappa_f \atop W} \left[ \sum_{i=1}^{N} \log_{\kappa_f} P_i(x_i \mid C_i)^{\otimes_{\kappa_f}^{1-\kappa_i}} \right]\right]^{1-\kappa_o} \tag{14}$$

If $\kappa_f = 0$ the coupled power term reduces to the standard power term $P_i^{1-\kappa_i}$ which is the expression for the coupled-probability. Likewise, the output probability is modified by the coupled-probability term $1-\kappa_o$.

## 4. Application to Designing and Assessing a Fusion Algorithm

The relationship between risk, generalized entropy, and the generalized means provides the foundation for a new method to characterize the performance of a fusion algorithm relative to requirements for robustness, accuracy, and decisiveness in a fusion algorithm. A fusion algorithm designed to minimize the Shannon surprisal or equivalently maximize the geometric mean (*i.e.*, the generalized mean with $\kappa = 0$) provides a 'Neutral Probability'. Bounding this measure of 'Accuracy' are 'Decisive' and 'Robust' probabilities measured using positive and negative values of *κ*, respectively. While a variety of options are available in determining the value of *κ* to use, we choose as an example the values ±0.5 in part because $\kappa = 0.5$ has a similar cost function to the Brier or mean-square average. The exact relationship between the coupled-surprisal and the Brier score is:

$$S_B = \sum_i (1-p_{true,i})^2 = \sum_i \left(-\ln_{\kappa=1} p_{true,i}\right)^2 \tag{15}$$

The computational experiment we will use here is for the classification of handwritten digits from a collection of Dutch utility maps. The data set consists of six different feature sets extracted from the same source images. The feature sets are Fourier: 76 Fourier coefficients of the character shapes; Profiles: 216 profile correlations; KL-coef: 64 Karhunen-Loève coefficients; Pixel: 240 pixel averages in 2 × 3 windows; Zernike: 47 Zernike moments; Morph: 6 morphological features. The data is publicly available from the Machine Learning Repository under the name 'mfeat' [40]. A similar experiment with a larger number of fusion techniques and more limited evaluation criteria can be found in [41]. The data set consists of 2000 instances, with 200 for each of the 10 numeral classes. We have allocated 100 instances for training and 100 for testing for each of the classes. Figure 2a shows example digits reconstructed from the feature set consisting of sampled pixel values.

A Bayesian classifier is trained for each of the six feature sets. In training our Bayesian classifier, we have assumed all classes to have the same covariance matrix with a different mean vector for each class. This assumption will result in linear decision boundaries. The posterior outputs from each classifier are fused using the alpha-beta fusion equation. Although only one fusion technique is used, through varying the parameters we are able to reconstruct and thus compare our results to standard techniques including naïve Bayes, averaging, and log-averaging.

Figures 2b and 2c show the classification performance with and without the fusion algorithm. The performance of the six feature sets varies from 26 misclassifications for set 2 to 327 misclassifications for set 6. The classification performance using the generalized mean to fuse the feature sets is shown in Figure 2c. The generalized mean is effective in filtering the poor performance of the weaker feature sets (1, 5, and 6), but does not improve upon the best feature set (2). Positive values of alpha act as smoothing filters relative to the log-average or equivalently the geometric mean at $\alpha = 0$. The best performance for this example is achieved with $\alpha = 0.25$. Negative values of alpha accentuate differences, which beyond $\alpha < -0.5$ significantly degrades the classification performance. The beta parameter which models the degree of effective independence does not modify the relative values of the class probabilities and is not influential on the classification performance.

Nevertheless, since the output probabilities of a fusion algorithm may be an input for a higher-level fusion process, the accuracy and robustness of the probabilities are important criteria for assessing the fusion performance. The coupled-surprisal and equivalently the generalized mean provide a method to directly examine the decisiveness, accuracy, and robustness characteristics. Figure 3a shows the histogram of probability outputs for the alpha-beta fusion method optimized against the Shannon surprisal ($\alpha = 0.4, \beta = 0.6$) and three common fusion methods; naïve-Bayes ($\alpha = 0, \beta = 1$), which assumes independence between the inputs; log-averaging ($\alpha = 0, \beta = 0$), which assumes the inputs are correlated, but does not smooth errors; and averaging ($\alpha = 1, \beta = 0$), which assumes both correlation and error in the inputs. Figure 3b shows the risk profile for each of the fusion methods formed by the generalized mean of the true class probabilities *versus* the coupling parameter $\kappa$. The histogram for naïve-Bayes is skewed severely toward 0 or 1 and as shown in Figure 3b only performs well against the decisive metric ($\kappa = 0.5$). The performance drops dramatically near $\kappa = -0.2$, indicating a lack of robustness. Against the neutral metric ($\kappa = 0.0$) log-averaging out performs averaging and naïve-Bayes. Here the modeling of correlation provides improvement. Log-averaging continues to perform well against the robust metric ($\kappa = -0.5$) indicating that in this example there is not a high degree of errors in the different features sets. Thus the averaging method, which would be advantageous in a situation where errors need to be smoothed, is in this example unnecessary.

**Figure 2.** (a) Examples of the handwritten numerals used as a classification and inferencing problem. (b) Individual misclassification for the six feature sets. (c) Fusion of the feature sets using the generalized mean with alpha varied between −1 and 2.

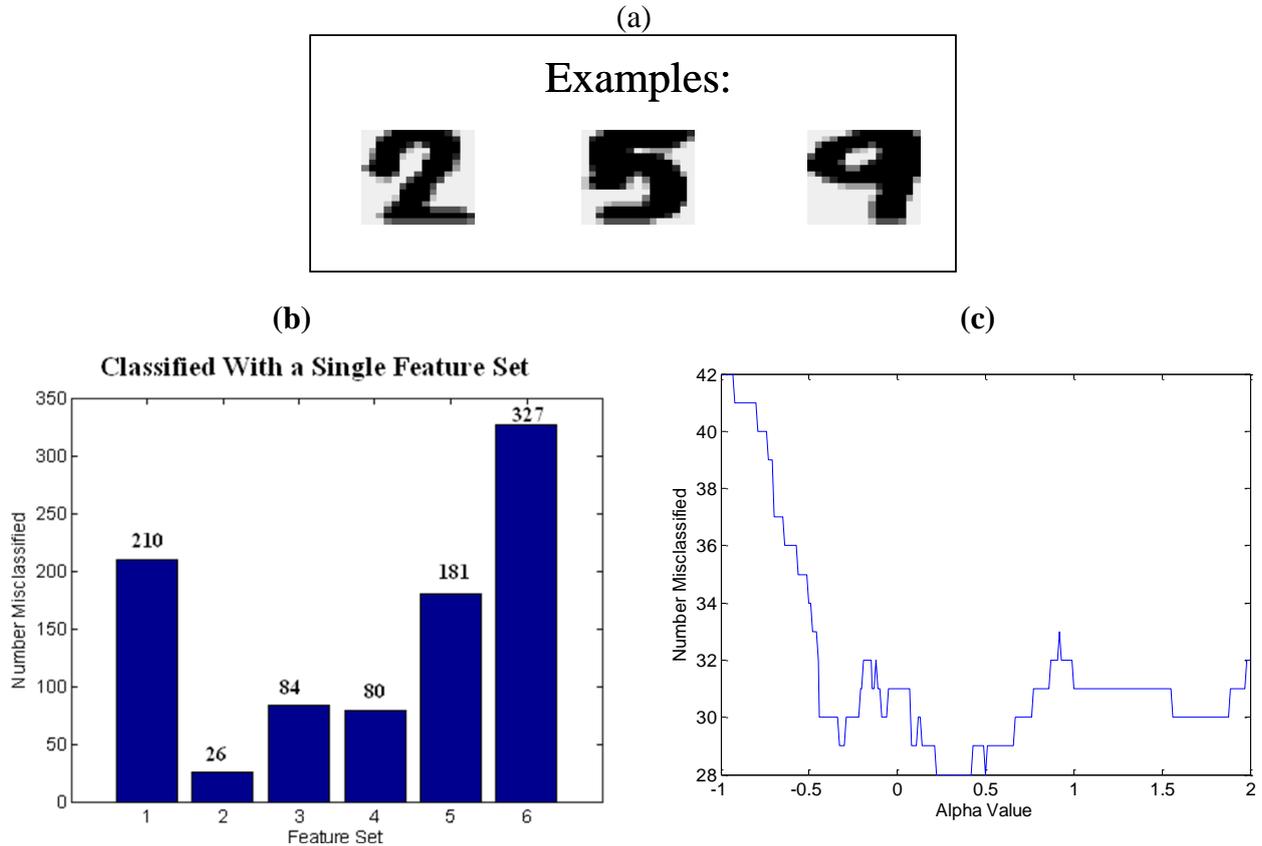

The fusion performance can be optimized using the alpha-beta algorithm described in Section 2.1. Figure 4 shows the performance of the alpha-beta algorithm against the Shannon, Brier, Decisive, and Robust metrics. Selection of the alpha-beta parameters could be based on one of these metrics or a combination of requirements. The optimal value for each metric is circled in each diagram. For Shannon surprisal the optimal value is $(\alpha = 0.4, \beta = 0.6)$ and its risk profile is shown in Figure 3b. This shows that the Decisive performance is very similar to the naïve Bayes and its Robust performance does not decay rapidly until $\kappa < -0.5$. Although the Brier as a proper score, provides an unbiased inference assessment, its origin in a linear $(\kappa = 1)$ metric, see Equation (15), makes it more favorable toward decisive algorithms, then the logarithmic metric $(\kappa = 0)$. In particular, forecasts extremely close to zero are not heavily penalized, which may be adequate in assessing the classification performance of an algorithm, but would lack robustness if the probability is an input to fusion algorithms utilizing Bayesian analysis. In this example, the optimal parameter for the effective independence moves from $\beta = 0.6$ to $0.8$, which will be more decisive and less robust.

**Figure 3.** (a) Histogram of the probabilities assigned to the true class for four fusion methods. (b) A risk profile based on the generalized mean of the true class probabilities *versus* the coupling parameter $\kappa$. The naïve-Bayes is a decisive fusion method which has near perfect score for large positive values of $\kappa$, but lacks robustness which is reflected in the sharp drop in the generalized mean for negative values of $\kappa$. Averaging and log-averaging are more robust methods; the generalized mean decays slower for negative values of $\kappa$ but does not achieve as high a value for positive $\kappa$. Using the alpha-beta fusion method, the Neutral (Shannon Surprisal) metric is optimized for $\alpha = 0.4, \beta = 0.6$ and has improved robustness relative to the naïve-Bayes.

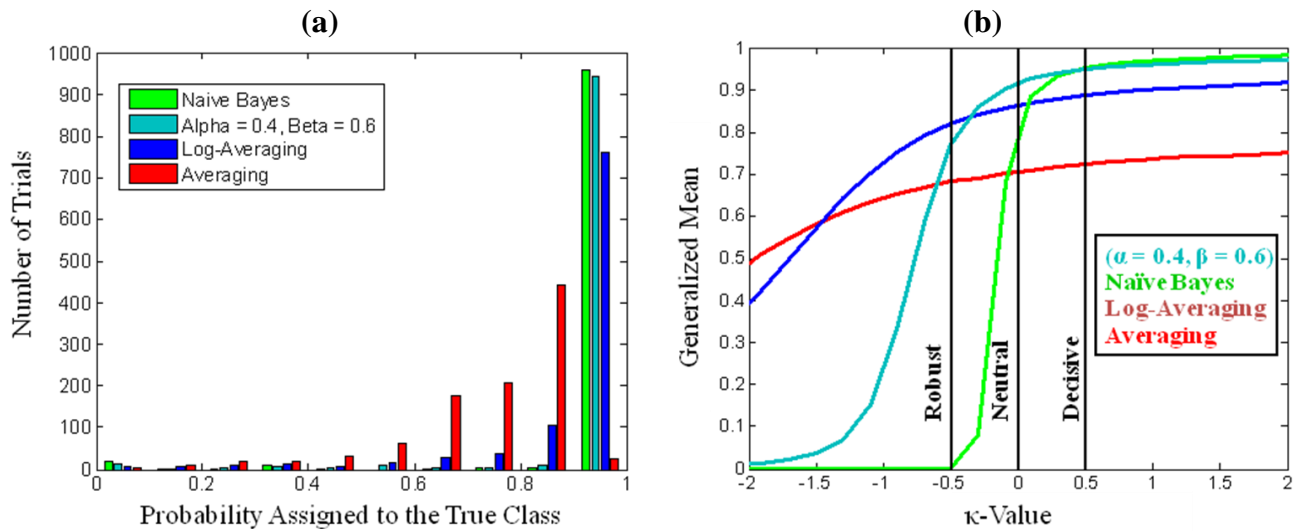

Algorithms like the Dempster-Shafer belief functions [42], in which probabilities based on Bayesian analysis are augmented by more decisive or more robust beliefs are anticipated. In the Dempster-Shafer methods, the Plausiblity and Belief functions are computed by considering power sets of the systems states. The power sets provide a means to consider the ambiguity in defining states, relaxing the assumption of mutually exclusive states, which is implicit in Bayesian analysis. In much the same way, the nonlinear statistical coupling method relaxes the assumption of mutually exclusive states by considering the coupled-probabilities of a system, which are nonlinear combinations of all the states. While the D-S methods provide a great deal of flexibility in how states are combined for particular beliefs, in practice simplifications are required to maintain computational effectiveness. While the nonlinear statistical coupling is a global mixing of the all states, the mathematical rigor of the approach provides both computational efficiency and analytical clarity. Furthermore, the weights $w_i$ of the alpha-beta fusion method in Equation (1) can be viewed as an individual risk bias on the inputs equal to $\kappa_i = 1 - w_i$. These algorithmic methods will be considered in more detail in a future publication.

**Figure 4.** Performance of alpha-beta fusion against effective probabilities based on the cost-function for (a) Shannon surprisal, κ = 0; (b) Brier or mean-square average; (c) Robust Coupled-Surprisal, κ = −0.5; and (d) Decisive Coupled-Surprisal, κ = 0.5. The circles indicate the region of optimal performance.

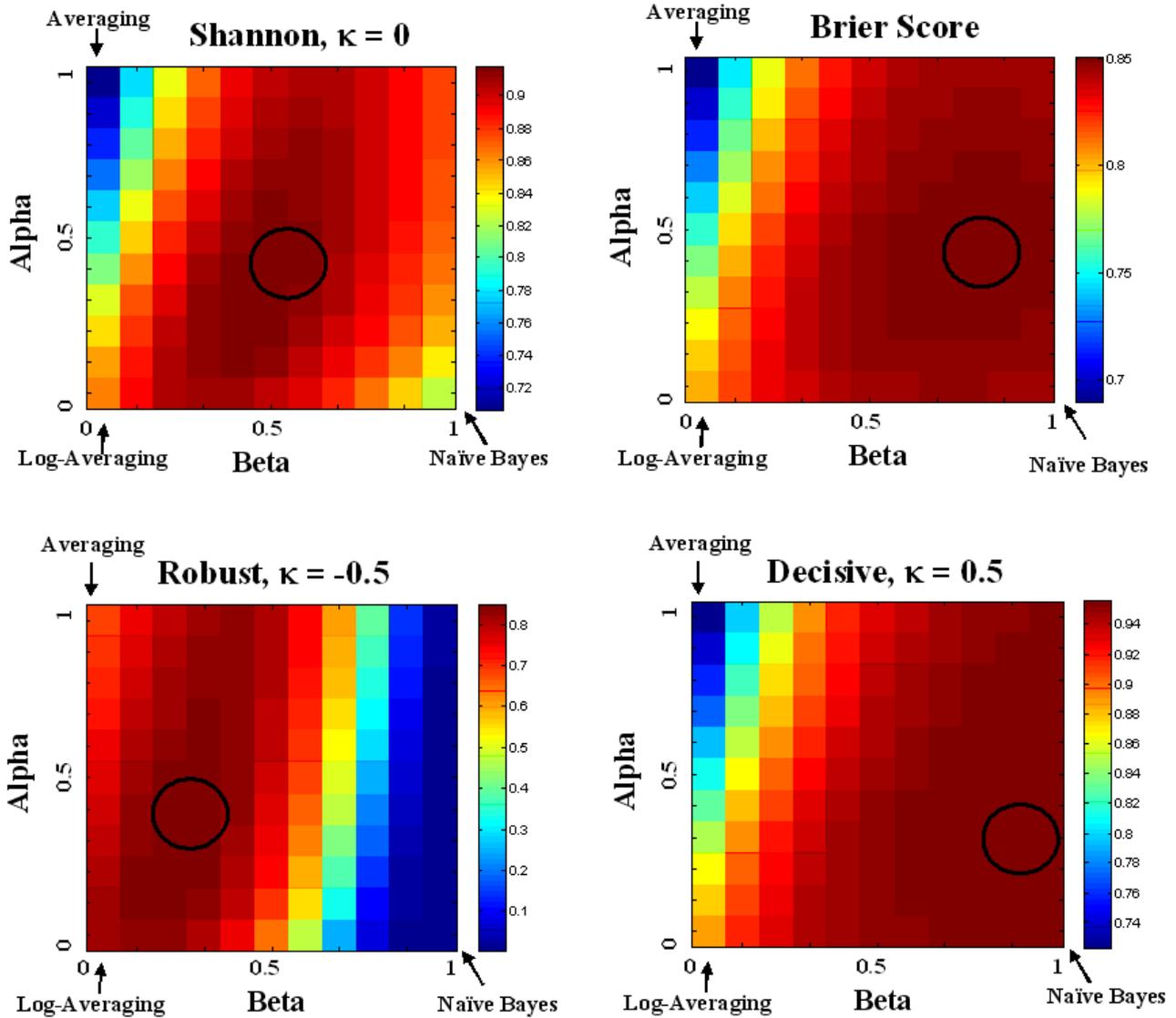

## 5. Conclusions

In this paper we have clarified the relationship between the generalized mean of probabilities and the generalized entropy functions defined by Tsallis and Renyi. The relationship is facilitated by defining as a physical property the nonlinear statistical coupling $\kappa$, which is a translation of the parameter $q$ defined by Tsallis and Renyi for the generalized entropy function, $\kappa = 1 - q$. Equations (8) and (9) show that both the Tsallis and Renyi entropy functions include the generalized mean of the probability states and there difference is the translation of the mean to an entropy scale using either the deformed logarithm for Tsallis entropy or the natural logarithm for Renyi entropy. Using the generalized mean of probabilities directly as a scoring rule provides useful intuition because the scale from 0 to 1 is simple to associate with the performance of a test.

The coupling parameter $\kappa$ of the generalized mean is associated with the degree of negative risk or optimism which biases the metric. The association with risk comes from the deformation of the accurate or neutral information cost function defined by the negative logarithm, known as Shannon surprisal. The performance of an inference algorithm is evaluated relative to the risk bias by measuring the average coupled-surprisal of the reported probabilities of the true class equation or equivalently the generalized mean of the true probabilities $P_{eff} = \left( \frac{1}{N} \sum_{i=1}^{N} p_{true,i}^{\kappa} \right)^{\frac{1}{\kappa}}$. As a starting point for exploring applications of this metric, we define Decisive ($\kappa = 0.5$), Neutral or Accurate ($\kappa = 0.0$), and Robust ($\kappa = -0.5$) probability metrics. Together these metrics highlight the balance between decisiveness and robustness for a fusion algorithm to be effective.

An effective fusion algorithm must model the potential errors and correlations between input nodes. Further the algorithm must be able to scale to large numbers of inputs without becoming computationally prohibitive. We have designed a fusion algorithm which uses the generalized mean to model smoothing of errors and a second parameter $N^{\beta}$ to model the effective number of independent samples given N inputs. Together the alpha-beta fusion algorithm:

$$P(C_i | X_1 = x_1, \ldots X_N = x_N) \approx \frac{1}{Z} \left( \frac{1}{N} \sum_{i=1}^{N} w_i P_i^{\alpha}(x_i | C_i) \right)^{\frac{N^{\beta}}{\alpha}} P(C_i)$$

provides a system-wide model of sensitivity and correlation. The alpha parameter is equivalent to the nonlinear statistical coupling $\kappa$. As the coupling between states is increased the output of the fusion is more uniform between the states resulting in a less decisive, but more robust inference. The beta parameter ranges from one effective sample ($\beta = 0$) to N effective samples ($\beta = 1$), providing a model of correlation between the samples.

## Acknowledgements

The authors have benefited from conversations with Ethan Phelps regarding proper scoring rules. Nora Tgavalekos is a co-inventor of the alpha-beta fusion algorithm. K. Nelson's collaborations with Sabir Umarov and Constantino Tsallis were influential in developing the concept of nonlinear statistical coupling. The authors were supported by Raytheon IRAD 10-IDS202.

## References and Notes

1. Dawid, A.P. The geometry of proper scoring rules. *Ann. Inst. Stat. Math.* **2007**, *59*, 77–93.
2. Gneiting, T.; Raftery, A.E. Strictly proper scoring rules, prediction, and estimation. *J. Am. Stat. Assoc.* **2007**, *102*, 359–378.
3. Jose, V.; Nau, R.F.; Winkler, R.L. Scoring rules, generalized entropy, and utility maximization. *Oper. Res.* **2008**, *56*, 1146.
4. Jaynes, E.T. *Probability Theory: The Logic of Science*; Cambridge University Press: Cambridge, UK, 2003.
5. Rosen, D.B. How good were those probability predictions? The expected recommendation loss (ERL) scoring rule. In Proceedings of the Thirteenth International Workshop on Maximum

Entropy and Bayesian Methods; Heidbreder, G.R., Ed.; Kluwer Academic Pub.: Santa Barbara, CA, USA, 1993; p. 401.

6. Wang, Q.A.; Nivanen, L.; Le Mehaute, A.; Pezeril, M. On the generalized entropy pseudoadditivity for complex systems. *J. Phys. A* **2002**, *35*, 7003–7007.
7. Furuichi, S.; Yanagi, K.; Kuriyama, K. Fundamental properties of Tsallis relative entropy. *J. Math. Phys.* **2004**, *45*, 4868.
8. Beck, C. Generalised information and entropy measures in physics. *Cont. Phys.* **2009**, *50*, 495–510.
9. Tsallis, C. Nonadditive entropy and nonextensive statistical mechanics-an overview after 20 years. *Braz. J. Phys.* **2009**, *39*, 337–356.
10. Nelson, K.P.; Umarov, S. Nonlinear statistical coupling. *Phys. A* **2010**, *389*, 2157–2163.
11. Borges, E.P. A possible deformed algebra and calculus inspired in nonextensive thermostatistics. *Phys. A* **2004**, *340*, 95–101.
12. Pennini, F.; Plastino, A.; Ferri, G.L. Fisher information, Borges operators, and q-calculus. *Phys. A* **2008**, *387*, 5778–5785.
13. Suyari, H.; Tsukada, M. Law of error in Tsallis statistics. *IEEE Trans. Inf. Theory* **2005**, *51*, 753–757.
14. Wada, T.; Suyari, H. κ-generalization of Gauss' law of error. *Phys. Lett. A* **2006**, *348*, 89–93.
15. Umarov, S.; Tsallis, C.; Steinberg, S. On a q-central limit theorem consistent with nonextensive statistical mechanics. *Milan J. Math.* **2008**, *76*, 307–328.
16. Kittler, J.; Hatef, M.; Duin, R.; Matas, J. On combining classifers. *IEEE Trans. Patt. Anal. Mach. Intel.* **1998**, *20*, 226.
17. Tax, D.; Van Breukelen, M.; Duin, R. Combining multiple classifiers by averaging or by multiplying? *Patt. Recognit.* **2000**, *33*, 1475–1485.
18. Kuncheva, L.I. *Combining Pattern Classifiers: Methods and Algorithms*; Wiley-Interscience: Hoboken, NJ, USA, 2004.
19. Hero, A.O.; Ma, B.; Michel, O.; Gorman, J. Alpha-divergence for classification, indexing and retrieval; *Technical Report CSPL-328*, U. Mich., Communication and Signal Processing Laboratory, May 2011.
20. Amari, S. Integration of stochastic models by minimizing α-divergence. *Neural Comp.* **2007**, *19*, 2780–2796.
21. Scannell, B.J.; McCann, C.; Nelson, K.P.; Tgavalekos, N.T. Fusion algorithm for the quantification of uncertainty in multi-look discrimination. Presented at the 8th Annual U.S. Missile Defense Conference, Washington, DC, USA, 22–24 March 2010.
22. Anteneodo, C.; Tsallis, C.; Martinez, A.S. Risk aversion in economic transactions. *Europhys. Lett.* **2002**, *59*, 635–641.
23. Anteneodo, C.; Tsallis, C. Risk aversion in financial decisions: A nonextensive approach. *arXiv* **2003**, arXiv:cond-mat/0306605v1.
24. Topsoe, F. On truth, belief and knowledge. In *ISIT'09*, Proceedings of the 2009 IEEE International Symposium on Information Theory, Seoul, Korea, 28 June–3 July 2009; Volume 1, pp. 139–143.
25. Tsallis, C. Possible generalization of Boltzmann-Gibbs statistics. *J. Stat. Phys.* **1988**, *52*, 479–487.

26. Gell-Mann, M.; Tsallis, C. *Nonextensive Entropy: Interdisciplinary Applications*; Oxford University Press: New York, NY, USA, 2004.
27. Vignat, C.; Plastino, A. Central limit theorem and deformed exponentials. *J. Phys. A* **2007**, *40*, F969–F978.
28. Marsh, J.A.; Fuentes, M.A.; Moyano, L.G.; Tsallis, C. Influence of global correlations on central limit theorems and entropic extensivity. *Phys. A* **2006**, *372*, 183–202.
29. Moyano, L.G.; Tsallis, C.; Gell-Mann, M. Numerical indications of a q-generalised central limit theorem. *Europhys. Lett.* **2006**, *73*, 813.
30. Hanel, R.; Thurner, S.; Tsallis, C.C. Limit distributions of scale-invariant probabilistic models of correlated random variables with the q-Gaussian as an explicit example. *Eur. Phys. J. B* **2009**, 72, 263–268.
31. Beck, C.; Cohen, E. Superstatistics. *Phys. A* **2003**, *322*, 267–275.
32. Wilk, G.; Wodarczyk, Z. Fluctuations, correlations and the nonextensivity. *Phys. A* **2007**, *376*, 279–288.
33. Nelson, K.P.; Umarov, S. The relationship between Tsallis statistics, the Fourier transform, and nonlinear coupling. *arXiv* **2008**, arXiv:0811.3777v1 [cs.IT].
34. Souto Martinez, A.; Silva González, R.; Lauri Espíndola, A. Generalized exponential function and discrete growth models. *Phys. A* **2009**, *388*, 2922–2930.
35. Kaniadakis, G.; Scarfone, A.M. A new one-parameter deformation of the exponential function. *Phys. A* 2002, 305, 69–75.
36. Kaniadakis, G.; Lissia, M.; Scarfone, A.M. Two-parameter deformations of logarithm, exponential, and entropy: A consistent framework for generalized statistical mechanics. *Phys. Rev. E* **2005**, 71, 46128.
37. Tsallis, C.; Plastino, A.R.; Alvarez-Estrada, R.F. Escort mean values and the characterization of power-law-decaying probability densities. *J. Math. Phys.* **2009**, *50*, 043303.
38. Abe, S. Stability of Tsallis entropy and instabilities of Renyi and normalized Tsallis entropies: A basis for q-exponential distributions. *Phys. Rev. E* **2002**, *66*, 46134.
39. Oikonomou, T. Tsallis, Renyi and nonextensive Gaussian entropy derived from the respective multinomial coefficients. *Phys. A* **2007**, *386*, 119–134.
40. Machine Learning Repository. Available online: http://www.ics.uci.edu/~mlearn/MLRepository.html (accessed on 15 November 2010).
41. Duin, R.; Tax, D. Experiments with classifier combining rules. In *Multiple Classifier Systems*, Springer: Berlin, Germany, 2000; Volume 1857, pp. 16–29.
42. Shafer, G. *A Mathematical Theory of Evidence*; Princeton University Press: Princeton, NJ, USA, 1976.